# Locally symmetric lattices for storage ring light sources

Zhenghe Bai[*], Penghui Yang, Guangyao Feng, Weimin Li, Lin Wang[†]

National Synchrotron Radiation Laboratory, University of Science and Technology of China, Hefei 230029, China

[*]baizhe@ustc.edu.cn
[†]wanglin@ustc.edu.cn

*Abstract*

In this paper, a new lattice concept called the locally symmetric lattice is proposed for storage ring light sources. In this new lattice, beta functions are made locally symmetric about two mirror planes of the lattice cell, and the phase advances between the two mirror planes satisfy the condition of nonlinear dynamics cancellation. There are two kinds of locally symmetric lattices, corresponding to two symmetric representations of lattice cell. In a locally symmetric lattice, main nonlinear effects caused by sextupoles can be effectively cancelled within one lattice cell, and generally there can also be many knobs of sextupoles available for further optimizing the nonlinear dynamics. Two kinds of locally symmetric lattices are designed for a 2.2 GeV diffraction-limited storage ring to demonstrate the lattice concept.

## I. INTRODUCTION

The fourth-generation storage ring light sources are being developed around the world with electron beam emittance reduced toward the diffraction-limited emittance of the emitted X-ray photons, which thus are also called diffraction-limited storage rings (DLSRs) [1]. To reduce the emittance within a moderate ring circumference, multi-bend achromat (MBA) lattices [2] are adopted in DLSR designs, and also unconventional bending magnets (bends), including longitudinal gradient bends (LGBs) [3-7] and reverse bends (RBs) [8-10], have been used in many DLSR lattices [11-16]. Compared with third-generation light sources, the MBA lattice of a DLSR generally creates larger natural chromaticity and lower dispersion, and thus stronger sextupoles are required, resulting in more serious nonlinear dynamics effects. To enlarge the dynamic aperture (DA) and momentum aperture (MA), nonlinear cancellation schemes are usually needed in DLSR lattice designs.

By choosing certain phase advances between sextupoles, many main nonlinear effects caused by sextupoles can be cancelled over a number of identical lattice cells or within one lattice cell [17]. The nonlinear cancellation of the PEP-X lattice design was done over 8 identical lattice cells [18], while the hybrid MBA lattice proposed by ESRF-EBS [11] and higher-order-achromat (HOA) lattice used by SLS-2 [19] are mainly based on the nonlinear cancellation within one lattice cell. Generally, the cancelation within one cell is more effective than that over a number of cells. The nonlinear cancellation in the hybrid MBA lattice is made by creating a pair of dispersion bumps separated by a -I transformation, and in the HOA lattice, a unit cell with certain phase advance is repeated a number of times. For the two cancellation strategies in the hybrid MBA and HOA lattices, the number of sextupole knobs is very limited in the ideal cancellation condition. More sextupole knobs are usually needed to mainly control tune shifts with amplitude and momentum, which are phase independent terms. To achieve desired nonlinear performance, as done in APS-U and SLS-2 lattice optimizations [20,13], one may have to organize sextupoles into more families at the cost of reduced periodicity or broken cancellation condition. As a contrast, in the PEP-X lattice with cancellation done over a number of lattice cells, many families of sextupoles can be used in the ideal cancellation condition. Thus, a question arises: can we have such lattices, in which not only the nonlinear cancellation is effectively done within one lattice cell as in the hybrid MBA and HOA lattice designs, but also many sextupole knobs can be used for further nonlinear optimization as in the PEP-X lattice design?

In this paper we will explore such lattices by making beta functions locally symmetric about two mirror planes of the lattice cell that are separated by certain phase advance. Under such a local symmetry, there is a translational invariance of phase advance that can be used for nonlinear cancellation. The concept of this new lattice that we call the locally symmetric lattice is described in Sec. II, which can be classified into two kinds. Then in Sec. III, locally symmetric lattices are designed for a DLSR with energy of 2.2 GeV, which is the same energy presently determined for Hefei Advanced Light Facility (HALF). Finally conclusion and outlook are given in Sec. IV. HALF is a new VUV and soft X-ray DLSR proposed by NSRL, and the HALF storage ring lattice is being designed with natural emittance goal of less than 100 pm·rad.

## II. LOCALLY SYMMETRIC LATTICE

### A. Basic lattice concept

In a basic lattice cell of a storage ring light source, the horizontal and vertical beta functions are generally symmetric about the midplane of the lattice cell. In the locally symmetric lattice, beta functions are further made locally symmetric about two mirror planes of the lattice cell, which have the same distance from the midplane, as shown in Fig. 1. In the locally symmetric regions, any position $s_i$ has a corresponding position $s_j$. For two corresponding positions, $s_i$ and $s_j$, the distance between them equals the distance between two mirror planes and they have the same Courant-Snyder parameters. Under such a local symmetry, from the relation between phase advance $\phi_u$ ($u = x, y$) and beta function $\beta_u$

$$\phi_u(s_1 \to s_2) = \int_{s_1}^{s_2} \frac{ds}{\beta_u(s)}, \qquad (1)$$

it follows that, as shown in Fig. 2, there exists a translational invariance: the horizontal and vertical phase advances between any two corresponding positions are the same, i.e.,

$$\phi_u(s_i \to s_j) = \phi_u(s_i + \Delta s \to s_j + \Delta s). \qquad (2)$$

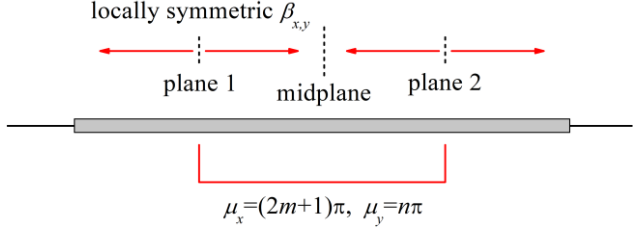

FIG. 1. Schematic of the locally symmetric lattice concept. The grey bar represents the magnet section of the lattice cell, and the two end parts are two halves of the straight section. The plane 1 and plane 2 are two mirror planes of the lattice.

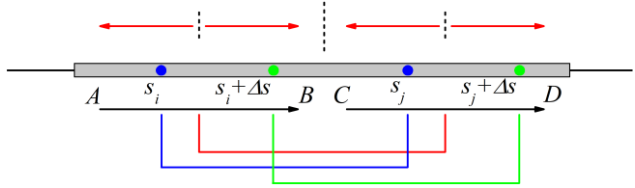

FIG. 2. Translational invariance of phase advance in the locally symmetric lattice. The phase advance between any position in the locally symmetric region [A, B] and its corresponding position in the other region [C, D] is the same.

The local symmetry of beta functions is the first condition of the locally symmetric lattice, and the second is that the horizontal and vertical phase advances between two mirror planes are set to

$$\phi_x = (2n_x+1)\pi, \quad \phi_y = n_y \pi, \qquad (3)$$

where $n_x$ and $n_y$ are integers. Thus, according to Eq. (2), the phase advances between any two corresponding positions also satisfy Eq. (3). The second condition is set for cancellation of the nonlinear effects caused by normal sextupoles [21], as we will show in the last part of this section. In a ring light source, skew sextupole fields are from manufacturing and alignment errors of magnets. For both normal and skew sextupole fields, the nonlinear cancellation condition Eq. (3) is changed to [21]

$$\phi_x = (2n_x+1)\pi, \quad \phi_y = (2n_y+1)\pi. \qquad (4)$$

With Eq. (4), the linear transfer map between any two corresponding positions is $-I$.

In a locally symmetric lattice, sextupoles are used in pairs and each pair of sextupoles with the same integrated strength are placed at two corresponding positions, so that many main nonlinear effects can be cancelled. There can be many different pairs of sextupoles in a locally symmetric lattice, and thus other nonlinear effects can be further minimized using these knobs so as to enlarge the DA and MA. In the actual lattice design, it is generally difficult and also not necessary to have perfect local symmetry of beta functions, and the local symmetry can be broken to a small extent.

### B. Locally symmetric lattice of the first kind

The locally symmetric lattices can be classified into two kinds according to the symmetric representations of a lattice cell, in which the midplane has different positions. In the usual symmetric representation of a lattice cell, where two halves of the straight section are on both sides of the cell and the arc section is in between them, the midplane is at the middle position of the arc section. The locally symmetric lattice in such a representation is called the locally symmetric lattice of the first kind, as shown in Fig. 3, where a schematic 8BA lattice is presented as an example.

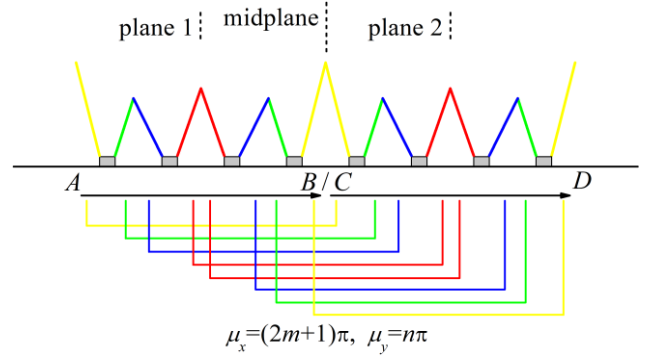

FIG. 3. Schematic of the locally symmetric lattice of the first kind, with an 8BA lattice as an example. The grey blocks represent main bends, and the upper color lines denote locally symmetric beta functions.

In this locally symmetric 8BA lattice, two mirror planes are at the middle position between the 2nd and 3rd bends and the middle position between the 6th and 7th bends. The locally symmetric region of the lattice is extended to the dispersion-free matching section, as indicated by the yellow lines, and correspondingly the dispersion between the 4th and 5th bends will also tend to be zero in the lattice design, which is unfavourable for emittance reduction. Employing RBs between the 4th and 5th bends can solve this problem, since RBs can disentangle the dispersion and beta functions [9]. Moreover, combining RBs with horizontally focusing quadrupoles can increase the horizontal damping partition, so that the emittance can be further reduced. Since the local symmetry is extended to the dispersion-free section, harmonic sextupoles can be also employed in the nonlinear optimization. We will design this 8BA lattice in the next section.

### C. Locally symmetric lattice of the second kind

There is also an uncommon symmetric representation of a lattice cell, where two halves of the arc section are on both sides of the cell and the straight section is in between them, and in this lattice representation, the midplane is at the middle position of the straight section. This representation can be transformed from the usual representation by a translation of half a lattice cell, expressed as

$$A_R B_L B_R A_L \xrightarrow{translation} B_R A_L A_R B_L, \quad (5)$$

where $A_L$ and $A_R$ denote the left and right halves of the straight section, respectively, and $B_L$ and $B_R$ the left and right halves of the arc section, respectively. The locally symmetric lattice in this uncommon representation is called the second kind. Fig. 4 shows an example of this kind, a schematic 6BA lattice. In this 6BA lattice, two mirror planes are at the middle position of the 2nd bend and the middle position of the 5th bend, and the local symmetry is also extended to the dispersion-free section, as indicated by the green lines. This 6BA lattice will be also designed later.

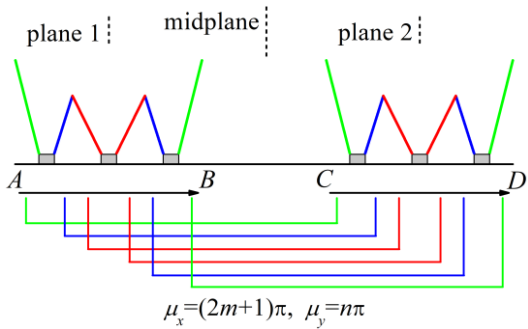

FIG. 4. Schematic of the locally symmetric lattice of the second kind, with a 6BA lattice as an example. The straight section is in the middle part of the cell.

## D. Nonlinear properties

For a section of storage ring lattice consisting of $n$ thin-lens sextupoles connected by $n+1$ linear transfer maps, based on the Lie algebra method, its transfer map can be written as [22]

$$M = A_0^{-1} e^{:f_3+f_4+\cdots:} R_{0,n+1} A_{n+1}, \quad (6)$$

where $A^{-1}$, the inverse of $A$, is a linear map for the normalization of the phase-space coordinates, $R_{0,n+1}$ is a rotational map with the angles being the betatron phase advances of the section, and $f_3$ and $f_4$ in the exponential Lie operator include many third- and fourth-order nonlinear terms. Generally, minimization of the nonlinear terms can improve the nonlinear dynamics performance, and lower-order terms are more important. For normal sextupoles, the third-order geometric terms in $f_3$ drive the resonances $\nu_x$, $3\nu_x$, $\nu_x-2\nu_y$ and $\nu_x+2\nu_y$, with the following form [22]

$$h_{jklm0} \propto \sum_{i=1}^{n} \lambda_i \beta_{x,i}^{(j+k)/2} \beta_{y,i}^{(l+m)/2} e^{i[(j-k)\mu_{x,i}+(l-m)\mu_{y,i}]}, \quad (7)$$

where the "0" in the subscript of $h_{jklm0}$ denotes geometric terms and $j+k+l+m=3$, and $\lambda_i$ is the integrated strength of the $i$-th sextupole.

In a locally symmetric lattice, sextupoles are used in pairs, and each pair of sextupoles have the same integrated strength and are placed at two corresponding positions with the same beta functions and the phase advances of Eq. (3). For each pair of sextupoles, all the third-order geometric resonance driving terms $h_{jklm0}$ [22] generated by themselves are cancelled. Therefore, due to that $h_{jklm0}$ in Eq. (7) is a sum over all sextupoles, all the third-order geometric terms are naturally cancelled within one cell of the locally symmetric lattice.

For the fourth-order geometric resonance driving terms, they can be minimized over a number of lattice cells by choosing certain cell phase advances [18]. For example, the horizontal and vertical phase advances of the ESRF-EBS lattice cell are near $(2+3/8, 7/8)\times 2\pi$, so that most of the fourth-order geometric resonances can be minimized over 8 lattice cells.

Besides, amplitude dependent tune shift terms ($\partial \nu_x / \partial J_x$, $\partial \nu_{x,y} / \partial J_{y,x}$ and $\partial \nu_y / \partial J_y$) and momentum dependent tune shift terms ($\partial^2 \nu_{x,y} / \partial \delta^2$ and $\partial^3 \nu_{x,y} / \partial \delta^3$) also need to be minimized in the nonlinear optimization. If the local symmetry is extended to a relatively large region, as will be designed in the next section, many families of sextupoles can be placed at corresponding positions in the locally symmetric lattice, and then the tune shift terms can be minimized using these available knobs. Generally, the tune shifts can be better controlled with more knobs of sextupoles. In addition, octupoles can be employed to help control the tune shifts [23]. With most of main geometric resonance terms being minimized in the locally symmetric lattice, the DA and MA can be enlarged by mainly minimizing the tune shift terms with the available knobs of sextupoles, which can be numerically optimized using evolutionary algorithms, such as genetic algorithm [24-27] and particle swarm optimization algorithm [28-30].

## III. LATTICE DESIGN

Now we will apply the locally symmetric lattice concept to the design of MBA lattices. Following the schematic lattices shown in Fig. 3 and Fig. 4, two kinds of locally symmetric lattices will be designed for a DLSR. The DLSR has an energy of 2.2 GeV and natural emittance goal of less than 100 pm·rad, which are the same as those of the HALF storage ring.

## A. Design of a locally symmetric 8BA lattice

According to the schematic in Fig. 3, we first design a locally symmetric 8BA lattice. The positions of two mirror planes of this lattice have been given in the previous section. In the design, the horizontal and vertical phase advances between the two mirror planes are set to $3\pi$ and $\pi$, respectively. To have more knobs of sextupoles, the region with locally symmetric beta functions includes the whole arc section and part of the dispersion-free matching section. One period of the designed 8BA lattice is shown in Fig. 5. In the lattice, all main bends are combined-function bends, and 3 families of combined-function RBs are employed to reduce the emittance. The straight section is 5.2 m long.

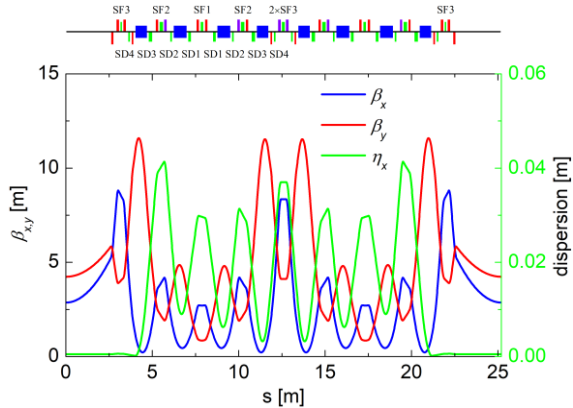

FIG. 5. Optical functions and magnet layout of the locally symmetric 8BA lattice. In the magnet layout, main bends are in blue, RBs in violet, quadrupoles in red and sextupoles in green.

To produce locally symmetric beta functions, the distribution of the quadrupole fields of magnets is almost locally symmetric about the two mirror planes in the lattice. For example, the 2$^{nd}$ and 3$^{rd}$ main bends have the same length and quadrupole field. But the distribution of the dipole fields is not locally symmetric, which changes the dispersion but has almost no effect on beta functions. So the 2$^{nd}$ and 3$^{rd}$ main bends have different dipole fields. To extend the local symmetry of beta functions to the dispersion-free section, an additional family of defocusing quadrupoles is added close to the 4$^{th}$ and 5$^{th}$ main bends, as shown in Fig. 5, since there are defocusing quadrupoles close to the 1$^{st}$ and 8$^{th}$ main bends in the dispersion-free section. And the RBs between the 4$^{th}$ and 5$^{th}$ main bends play an important role in emittance reduction, which help to increase the dispersion between these two bends.

The linear optics of the lattice was designed and optimized using a multi-objective particle swarm optimization (MOPSO) algorithm. In the optics optimization, the local symmetry of beta functions was an important constraint, and we introduced the relative beta function difference of two corresponding positions

$$\delta_\beta = \frac{|\beta_u(s_i) - \beta_u(s_j)|}{\beta_u(s_i)}, \quad (8)$$

which was set to less than several percent for the locally symmetric region in the constraint. Smaller $\delta_\beta$ gives better local symmetry of beta functions. It can be seen from Fig. 5 that the designed 8BA lattice has almost perfect local symmetry of beta functions. Table I lists the main parameters of the designed storage ring. The natural emittance is 53.6 pm·rad. The betatron tunes are situated on a linear coupling resonance for producing round beams. The horizontal and vertical phase advances of one lattice cell are near (3+1/4, 1+1/4)×2π so as to minimize the third- and fourth-order geometric resonances over 4 lattice cells.

TABLE I. Main parameters of the storage ring designed with the locally symmetric 8BA lattice.

| Parameter | Value |
|---|---|
| Energy ($E$) | 2.2 GeV |
| Circumference ($C$) | 403.2 m |
| Number of periods ($N_p$) | 16 |
| Natural emittance ($\varepsilon_{x,0}$) | 53.6 pm·rad |
| Betatron tunes ($v_x/v_y$) | 52.28/19.28 |
| Natural chromaticities ($\xi_x/\xi_y$) | -91/-56 |
| Natural damping times ($\tau_x/\tau_y/\tau_z$) | 16/30.9/28.8 ms |
| Damping partitions ($J_x/J_y/J_z$) | 1.93/1.0/1.07 |
| Momentum compaction factor ($\alpha$) | 7.3×10$^{-5}$ |
| Natural energy spread ($\sigma_\delta$) | 0.73×10$^{-3}$ |
| $\beta_x$, $\beta_y$ at the middle of straight sections ($\beta_{x,0}/\beta_{y,0}$) | 2.87/4.24 m |

There are 7 families of sextupoles in the lattice, including 3 focusing and 4 defocusing ones, which are all located in the locally symmetric region, as shown in the top of Fig. 5. The arrangement of sextupoles is locally symmetric about the two mirror planes, so that it is also symmetric about the midplane of the cell. Note that as shown in Fig. 5, there are two sextupoles of the same family (2×SF3) in the middle of the cell, which can be combined into one sextupole with twice the integrated strength of SF3. Half of sextupoles of the families SD4 and SF3 are located in the dispersion-free section, acting as harmonic sextupoles with no contribution to chromaticity correction. The nonlinear dynamics was optimized with these 7 families of sextupoles using MOPSO, and the horizontal and vertical chromaticities were corrected to (2, 2). In the optimization, the objectives were to maximize the effective DA and to minimize the momentum dependent tune shifts. The effective DA refers to the area without amplitude dependent tunes across integer and half-integer resonances.

Fig. 6 shows the frequency map analysis of the optimized DA, which was tracked for 1024 turns. It is seen that the horizontal DA is larger than 3 mm and the amplitude dependent tune shifts are moderate. A small DA area with y>3 mm is not shown in the figure, which has larger tune excursion but is not the main DA area. Fig. 7 shows the momentum dependent tunes, which are controlled in relatively small ranges for relative momentum deviations of -5%~5%. According to the formulae of the third-order geometric resonance driving terms [22], we calculated the change of these terms along one lattice cell, which is shown in Fig. 8. We see that all the third-order geometric terms are almost cancelled within one cell. The residual values are due to very small deviation from the perfect local symmetry in the lattice design.

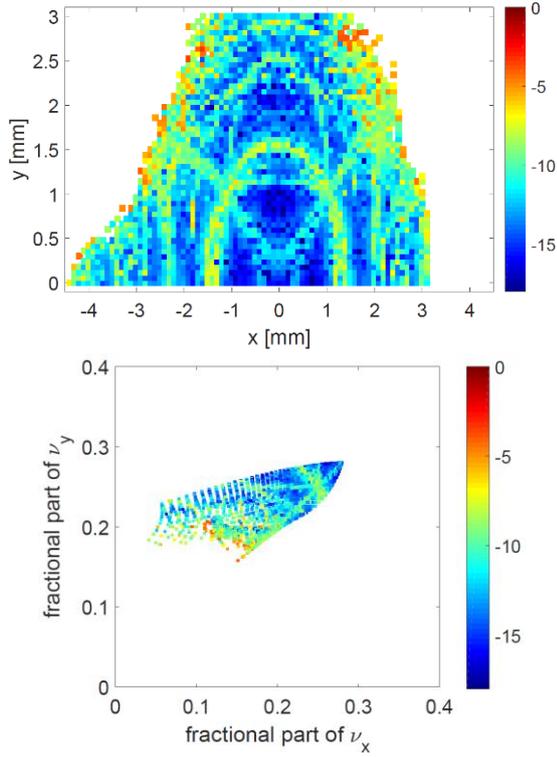

FIG. 6. Frequency map analysis for the DA of the locally symmetric 8BA lattice, tracked at the middle of the straight section. The color bar denotes the tune diffusion, with blue for more regular motion and red for chaotic.

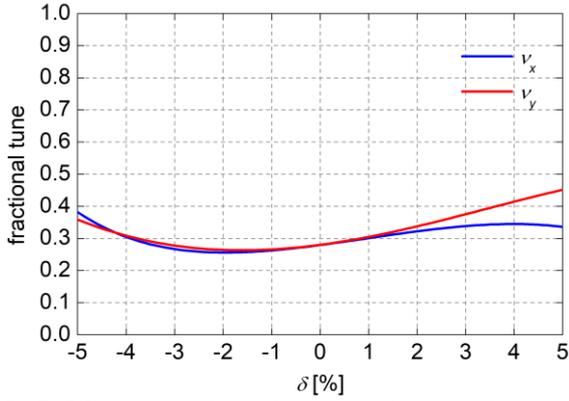

FIG. 7. Momentum dependent tune shifts of the locally symmetric 8BA lattice.

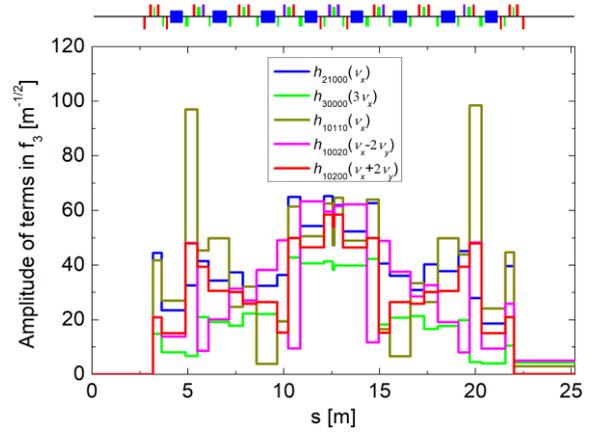

FIG. 8. Change of third-order geometric resonance driving terms along one cell of the locally symmetric 8BA lattice.

## B. Design of locally symmetric 6BA lattices

Next, according to the schematic in Fig. 4, we design two locally symmetric 6BA lattices, one with almost perfect local symmetry as designed for the 8BA lattice and the other with broken local symmetry. For the two 6BA lattices, in the usual symmetric representation of lattice, two mirror planes are at the middle position of the 5$^{th}$ main bend of the previous lattice cell and the middle position of the 2$^{nd}$ main bend of the current cell, and the horizontal and vertical phase advances between them are also $3\pi$ and $\pi$, respectively. The locally symmetric region is also extended to the dispersion-free matching section in the design to have more sextupole knobs for nonlinear optimization. As in the 8BA lattice, MOPSO was used to optimize the linear optics and nonlinear dynamics of the 6BA lattices.

One period of the designed 6BA lattice with almost perfect local symmetry is shown in Fig. 9, which has 6 combined-function main bends and 3 families of combined-function RBs. The straight section is 5.2 m long. The main parameters of the storage ring are listed in Table II. The natural emittance is 70.9 pm·rad. The third- and fourth-order geometric resonances are also considered to be minimized over 4 lattice cells by setting the horizontal and vertical phase advances of one cell near $(2+3/4, 3/4)\times 2\pi$. As shown in Fig. 9, there are 5 families of sextupoles used for nonlinear optimization.

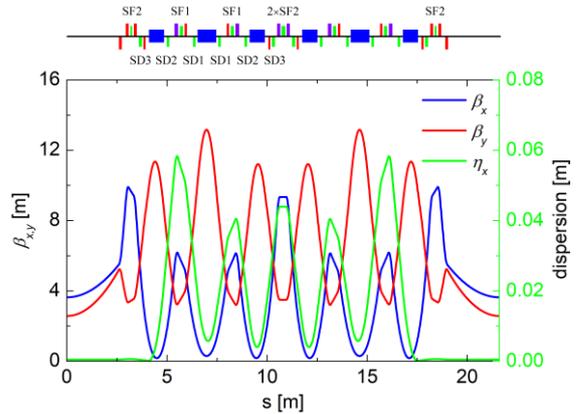

FIG. 9. Optical functions and magnet layout of the 6BA lattice with almost perfect local symmetry.

TABLE II. Main storage ring parameters of two locally symmetric 6BA lattices with almost perfect symmetry (Case A) and with broken symmetry (Case B).

| Parameter | Case A | Case B |
|---|---|---|
| $E$ | 2.2 GeV | |
| $C$ | 345.6 m | |
| $N_p$ | 16 | |
| $\varepsilon_{x,0}$ | 70.9 pm·rad | 57.4 pm·rad |
| $\nu_x/\nu_y$ | 44.22/11.22 | 44.28/11.28 |
| $\xi_x/\xi_y$ | -101/-48 | -107/-46 |
| $\tau_x/\tau_y/\tau_z$ | 12.5/23.3/20.4 ms | 12.6/24.4/22.9 ms |
| $J_x/J_y/J_z$ | 1.86/1.0/1.14 | 1.93/1.0/1.07 |
| $\alpha$ | $7.0\times 10^{-5}$ | $8.0\times 10^{-5}$ |
| $\sigma_\delta$ | $0.75\times 10^{-3}$ | $0.81\times 10^{-3}$ |
| $\beta_{x,0}/\beta_{y,0}$ | 3.64/2.58 m | 4.41/2.43 m |

With the horizontal and vertical chromaticities corrected to (2, 2), the optimized on- and off-momentum DAs are shown in Fig. 10, and the tune shifts with amplitude and momentum are shown in Fig. 11. We see that the on-momentum horizontal DA is about 4 mm and, interestingly, off-momentum DAs are even larger than the on-momentum DA. The momentum dependent horizontal tune approaches the half-integer resonance at the relative momentum deviation of 4%. Compared to the 8BA lattice with more sextupole knobs, the momentum dependent tune shift of this 6BA lattice is relatively larger. As shown in Fig. 12, all the third-order geometric resonance driving terms are almost cancelled within one cell of this 6BA lattice. Note that these terms were calculated from the middle position of the arc section (i.e., in the uncommon symmetric representation of lattice).

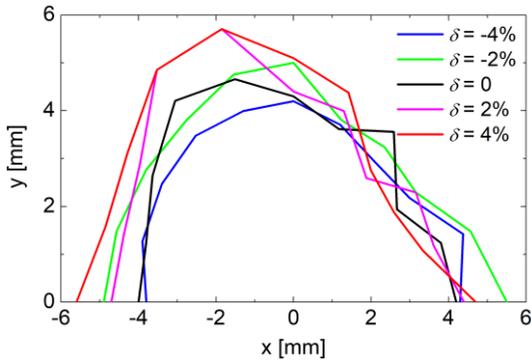
FIG. 10. On- and off- momentum DAs of the 6BA lattice with almost perfect local symmetry, tracked at the middle of the straight section.

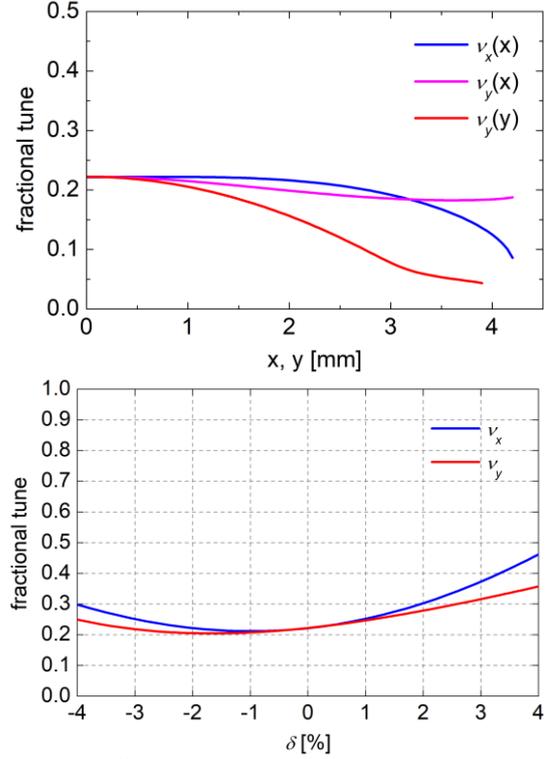
FIG. 11. Amplitude and momentum dependent tune shifts of the 6BA lattice with almost perfect local symmetry.

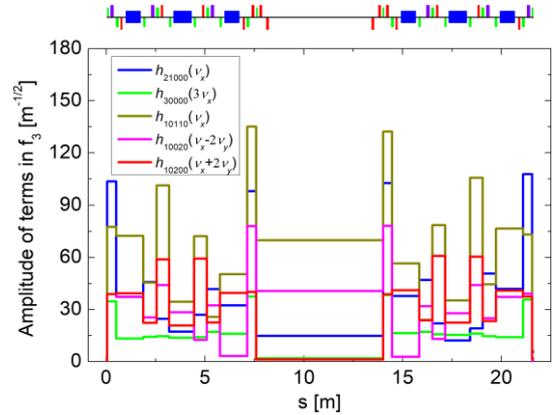
FIG. 12. Change of third-order geometric resonance driving terms along one cell of the 6BA lattice with almost perfect local symmetry.

Further numerical study showed that for this 6BA lattice, it is not necessary to make almost perfect local symmetry and better nonlinear dynamics performance can be achieved by both breaking the local symmetry to a slightly larger extent and grouping the sextupoles into more families. A 6BA lattice with broken local symmetry was designed and is shown in Fig. 13. For reducing the emittance, the 3rd and 4th main bends of the lattice are LGBs, with defocusing quadrupoles adjacent to them. In the linear optics optimization, $\delta_\beta$ was constrained to be less than 15%. Note that the difference between the beta functions in the 1st and 3rd main bends was not constrained. The LGBs are of the same distribution of dipole field, and each has 7 pieces with a symmetric field distribution. The middle piece has the highest dipole field of about 1.45 T, which can be used as the radiation source of bend beamline. The natural emittance was reduced to 57.4 pm·rad, and other main parameters are shown in Table II. Since the local symmetry was broken to some extent, sextupoles at two corresponding positions can have different strengths. There are 8 families of sextupoles, as shown in Fig. 13. In the nonlinear optimization, the horizontal and vertical chromaticities were corrected to (2, 2). The optimized on- and off-momentum DAs are shown in Fig. 14, and the tune shifts with amplitude and momentum are shown in Fig. 15. It is seen that the on-momentum horizontal DA is enlarged to about 6 mm and the horizontal tune shift with momentum is better controlled. In this paper, we have not used octupoles in the lattices, which can be considered as extra knobs for further nonlinear optimization.

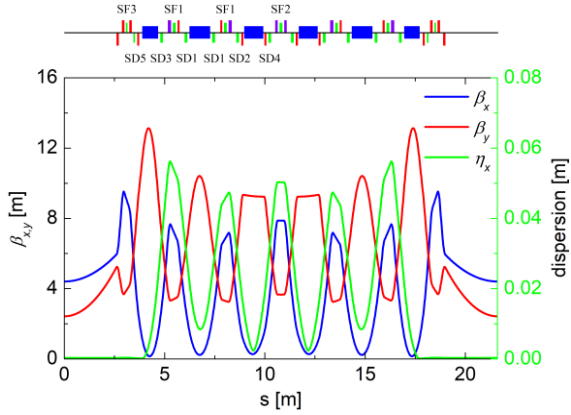

FIG. 13. Optical functions and magnet layout of the 6BA lattice with broken local symmetry.

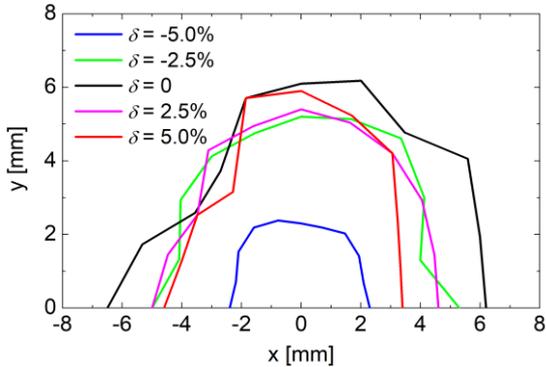

FIG. 14. On- and off- momentum DAs of the 6BA lattice with broken local symmetry, tracked at the middle of the straight section.

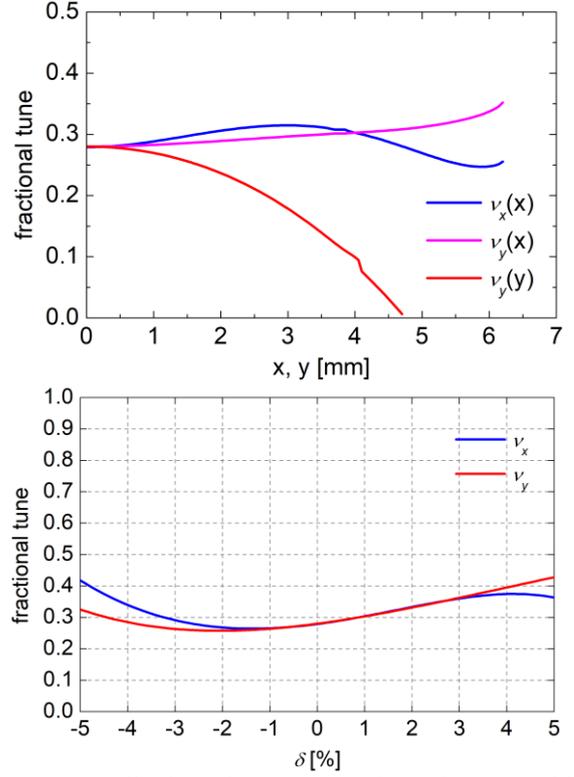

FIG. 15. Amplitude and momentum dependent tune shifts of the 6BA lattice with broken local symmetry.

Compared with the 8BA lattice, the 6BA lattice with broken local symmetry has slightly larger natural emittance. However, when the intra-beam scattering (IBS) is considered, the 6BA lattice has lower equilibrium emittance than the 8BA lattice due to relatively shorter damping time. For a round beam with current of 500 mA and bunches lengthened by a factor of 3 (500 MHz RF cavity, 80% of RF buckets equally filled), the equilibrium emittance, $\varepsilon_x = \varepsilon_y$, of the 6BA lattice is 57 pm·rad (increased by 50% compared to the zero-current emittance of 38 pm·rad), and the equilibrium emittance of the 8BA lattice is 66 pm·rad. The IBS-induced emittance growth will be suppressed when damping wigglers and insertion devices are added, which produce additional radiation damping. Moreover, compared with the 8BA lattice, the storage ring designed with the 6BA lattice has shorter circumference. Therefore the 6BA lattice is a better lattice candidate. The Touschek lifetime of the 6BA lattice is more than 40 hours for a round beam with current of 500 mA (without including errors and bunch lengthening), which was calculated based on the local MA shown in Fig. 16. The asymmetry of the local MA at the straight section is due to the distortion of RF bucket. Swap-out injection [31] can be considered as the injection scheme for the 6BA lattice. Besides the MBA lattices presented in this paper, lattices based on the hybrid MBA concept are also being designed and optimized for HALF, which have 20 periods and larger DAs that can support the implementation of off-axis injection [32, 30].

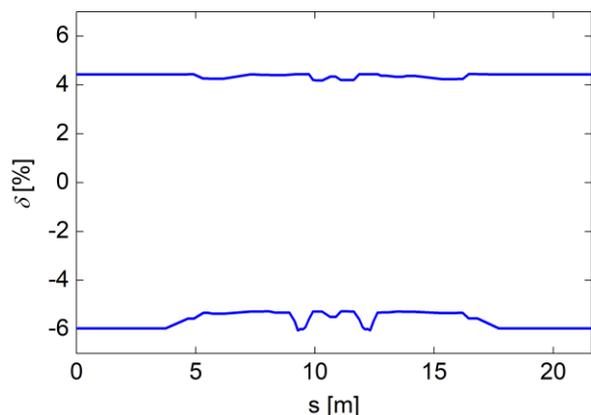

FIG. 16. Local MA from 6D tracking for one period of the 6BA lattice with broken local symmetry.

## IV. CONCLUSION AND OUTLOOK

A new lattice concept, the locally symmetric lattice, was proposed for designing storage ring light sources. In this new lattice, the local symmetry of beta functions provides a translational invariance of phase advances. With each pair of identical sextupoles located at any two corresponding positions, all the third-order resonance driving terms can be effectively cancelled within one lattice cell, and generally, there can be many available knobs of sextupoles for further minimizing other nonlinear terms, including amplitude and momentum dependent tune shifts. Two kinds of locally symmetric lattices have been designed with natural emittances of less than 100 pm·rad for HALF, and their nonlinear dynamics performances demonstrated the nonlinear properties of the locally symmetric lattices. One of the designed lattices, the 6BA lattice with local symmetry broken to some extent, presented lower equilibrium emittance and better nonlinear dynamics performance, and thus can be considered as a better lattice candidate for HALF.

We can clearly see from the designed lattices that a locally symmetric lattice has two quasi-periodic parts. Therefore, the locally symmetric lattice concept can be naturally extended to a super-period case with two basic lattice cells. Fig. 17 shows the schematic of such a super-period locally symmetric lattice concept. In this super-period lattice, the midplane is at the middle position of one straight section, and two mirror planes are at the middle positions of the arc sections of two basic cells. The super-period lattice can have straight sections with alternating high and low beta functions, and high-beta straight section is beneficial for injection and low-beta one for enhancing the insertion device brightness. Actually, the locally symmetric 6BA lattice can be changed into a super-period locally symmetric triple-bend-achromat lattice.

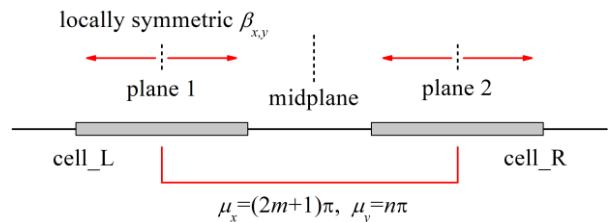

FIG. 17. Schematic of the super-period locally symmetric lattice with two basic cells.

## ACKNOWLEDGMENTS

This work was supported by the National Natural Science Foundation of China (No. 11875259) and the National Key Research and Development Program of China (No. 2016YFA0402000).

[1] R. Hettel, DLSR design and plans: an international overview, J. Synchrotron Radiat. **21**, 843 (2014).
[2] D. Einfeld, M. Plesko, and J. Schaper, First multi-bend achromat lattice consideration, J. Synchrotron Radiat. **21**, 856 (2014).
[3] J. Guo and T. Raubenheimer, Low emittance e-/e+ storage ring design using bending magnets with longitudinal gradient, in *Proceedings of the 8th European Particle Accelerator Conference, Paris, 2002* (EPS-IGA and CERN, Geneva, 2002), p. 1136.
[4] Y. Papaphilippou and P. Elleaume, Analytical considerations for reducing the effective emittance with variable dipole field strengths, in *Proceedings of the 21st Particle Accelerator Conference, Knoxville, TN, 2005* (IEEE, Piscataway, NJ, 2005), p. 2086.
[5] R. Nagaoka and A. F. Wrulich, Emittance minimisation with longitudinal dipole field variation, Nucl. Instrum. Methods Phys. Res., Sect. A **575**, 292 (2007).
[6] C.-x. Wang, Minimum emittance in storage rings with uniform or nonuniform dipoles, Phys. Rev. Accel. Beams **12**, 061001 (2009).
[7] A. Streun and A. Wrulich, Compact low emittance light sources based on longitudinal gradient bending magnets, Nucl. Instrum. Methods Phys. Res., Sect. A **770**, 98 (2015).
[8] J. Delahaye and J. Potier, Reverse bending magnets in a combined-function lattice for the CLIC damping ring, in *Proceedings of the 1989 Particle Accelerator Conference, Chicago, IL* (IEEE, New York, 1989), p. 1611.
[9] A. Streun, The anti-bend cell for ultralow emittance storage ring lattices, Nucl. Instrum. Methods Phys. Res., Sect. A **737**, 148 (2014).
[10] B. Riemann and A. Streun, Low emittance lattice design from first principles: Reverse bending and longitudinal gradient bends, Phys. Rev. Accel. Beams **22**, 021601 (2019).
[11] L. Farvacque, N. Carmignani, J. Chavanne, A. Franchi, G. L. Bec, S. Liuzzo, B. Nash, T. Perron, and P. Raimondi, A low-emittance lattice for the ESRF, in *Proceedings of the 4th International Particle Accelerator*

Conference (IPAC2013), Shanghai, China, 2013* (JA-CoW, Geneva, 2013), p. 79.

[12] M. Borland, Y. Sun, V. Sajaev, R.R. Lindberg, and T. Berenc, Lower emittance lattice for the Advanced Photon Source upgrade using reverse bending magnets, in *Proceedings of NAPAC2016, Chicago, IL, USA, 2016* (JACoW, Geneva, 2017), p. 877.

[13] A. Streun, T. Garvey, L. Rivkin, V. Schlott, T. Schmidt, P. Willmott, and A. Wrulich, SLS-2—the upgrade of the Swiss Light Source, J. Synchrotron Radiat. **25**, 631 (2018).

[14] Y. Jiao, G. Xu, X. H. Cui et al., The HEPS project, J. Synchrotron Radiat. **25**, 1611 (2018).

[15] E. Karantzoulis, A. Carniel, S. Krecic, Elettra, present and future, in *Proceedings of the 10th International Particle Accelerator Conference (IPAC2019), Melbourne, Australia, 2019* (JACoW, Geneva, 2019), p. 1468.

[16] P. Yang, W. Li, Z. Ren, Z. Bai, and L. Wang, Design of a diffraction-limited storage ring lattice using longitudinal gradient bends and reverse bends, Nucl. Instrum. Methods Phys. Res., Sect. A **990**, 164968 (2021).

[17] M. Borland, G. Decker, L. Emery, V. Sajaev, Y. Sun, and A. Xiao, Lattice design challenges for fourth-generation storage-ring light sources, J. Synchrotron Radiat. **21**, 912 (2014).

[18] Y. Cai, K. Bane, R. Hettel, Y. Nosochkov, M.-H. Wang, and M. Borland, Ultimate storage ring based on fourth-order geometric achromats, Phys. Rev. Accel. Beams **15**, 054002 (2012).

[19] J. Bengtsson and A. Streun, Robust Design Strategy for SLS-2, Paul Scherrer Institut Technical Report No. SLS2-BJ84-001-2, 2017.

[20] M. Borland, M. Abliz, N. Arnold et al., The upgrade of the Advanced Photon Source, in *Proceedings of the 9th International Particle Accelerator Conference (IPAC2018), Vancouver, BC, Canada, 2018* (JACoW, Geneva, 2018), p. 2872.

[21] R. Bartolini, Design and optimization strategies of nonlinear dynamics in diffraction-limited synchrotron light sources, in *Proceedings of the 7th International Particle Accelerator Conference (IPAC2016), Busan, Korea, 2016* (JACoW, Geneva, 2016), p. 33.

[22] J. Bengtsson, The sextupole scheme for the Swiss Light Source (SLS): An analytic approach, SLS Note No. 9/97, 1997.

[23] S. C. Leemann and A. Streun, Perspectives for future light source lattices incorporating yet uncommon magnets, Phys. Rev. Accel. Beams **14**, 030701 (2011).

[24] M. Borland, V. Sajaev, L. Emery, and A. Xiao, Direct methods of optimization of storage ring dynamic and momentum aperture, in *Proceedings of the 23rd Particle Accelerator Conference, Vancouver, Canada, 2009* (IEEE, Piscataway, NJ, 2009), p. 3850.

[25] L. Yang, Y. Li, W. Guo, and S. Krinsky, Multiobjective optimization of dynamic aperture, Phys. Rev. Accel. Beams **14**, 054001 (2011).

[26] M. P. Ehrlichman, Genetic algorithm for chromaticity correction in diffraction limited storage rings, Phys. Rev. Accel. Beams **19**, 044001 (2016).

[27] Y. Li, W. Cheng, L. H. Yu, and R. Rainer, Genetic algorithm enhanced by machine learning in dynamic aperture optimization, Phys. Rev. Accel. Beams **21**, 054601 (2018).

[28] Z. Bai, L. Wang, and W. Li, Enlarging dynamic and momentum aperture by particle swarm optimization, in *Proceedings of the 2nd International Particle Accelerator Conference (IPAC2011), San Sebastián, Spain, 2011* (EPS-AG, Spain, 2011), p. 948.

[29] X. Huang and J. Safranek, Nonlinear dynamics optimization with particle swarm and genetic algorithms for SPEAR3 emittance upgrade, Nucl. Instrum. Methods Phys. Res., Sect. A **757**, 48 (2014).

[30] J. Xu, P. Yang, G. Liu, Z. Bai, and W. Li, Constraint handling in constrained optimization of a storage ring multi-bend-achromat lattice, Nucl. Instrum. Methods Phys. Res., Sect. A **988**, 164890 (2021).

[31] L. Emery and M. Borland, Possible long-term improvements to the Advanced Photon Source, in *Proceedings of the 2003 Particle Accelerator Conference, Portland, Oregon, 2003* (IEEE, New York, 2003), p. 256..

[32] Z. Bai, Internal report at NSRL, 2020.